\documentclass[pdflatex,sn-apa]{sn-jnl}

\usepackage{graphicx}%
\usepackage{multirow}%
\usepackage{amsmath,amssymb,amsfonts}%
\usepackage{amsthm}%
\usepackage{mathrsfs}%
\usepackage[title]{appendix}%
\usepackage{xcolor}%
\usepackage{textcomp}%
\usepackage{manyfoot}%
\usepackage{booktabs}%
\usepackage{algorithm}%
\usepackage{algorithmicx}%
\usepackage{algpseudocode}%
\usepackage{listings}%

\theoremstyle{thmstyleone}%
\theoremstyle{thmstyletwo}%

\theoremstyle{thmstylethree}%

\unnumbered 

\begin{document}

\title[Neural Earthquake Forecasting with Minimal Information: Limits, Interpretability, and the Role of Markov Structure]{Neural Earthquake Forecasting with Minimal Information: Limits, Interpretability, and the Role of Markov Structure}


\author[1, 2]{\fnm{Jonas} \sur{Köhler}}\email{jkoehler@fias.uni-frankfurt.de}

\author*[1,2]{\fnm{Nishtha} \sur{Srivastava}}\email{N.Srivastava@em.uni-frankfurt.de}
\equalcont{These authors contributed equally to this work.}

\author[3]{\fnm{Kai} \sur{Zhou}}\email{zhoukai@cuhk.edu.cn}
\equalcont{These authors contributed equally to this work.}

\author[1]{\fnm{Claudia} \sur{Quinteros}}\email{quinteros@fias.uni-frankfurt.de}
\equalcont{These authors contributed equally to this work.}

\author[1,4]{\fnm{Johannes} \sur{Faber}}\email{FAIS.research@proton.me}
\equalcont{These authors contributed equally to this work.}

\author[5]{\fnm{F.~Alejandro} \sur{Nava}}\email{fnava@cicese.mx}
\equalcont{These authors contributed equally to this work.}

\affil*[1]{\orgname{Frankfurt Institute of Advanced Studies}, \orgaddress{\street{Ruth-Moufang-Str.~1}, \city{Frankfurt am Main}, \postcode{60438}, \state{Hesse}, \country{Germany}}}

\affil[2]{\orgdiv{Institute of Geosciences}, \orgname{Goethe-University Frankfurt}, \orgaddress{\street{Altenhöferallee 1}, \city{Frankfurt am Main}, \postcode{60438}, \state{Hesse}, \country{Germany}}}

\affil[3]{\orgdiv{Department}, \orgname{The Chinese University of Hong Kong, Shenzhen}, \orgaddress{ \city{Shenzhen}, \postcode{518172}, \state{Guangdong}, \country{China}}}

\affil[4]{\orgdiv{Institute for Theoretical Physics}, \orgname{Goethe-University Frankfurt}, \orgaddress{\street{Max-von-Laue-Str.~1}, \city{Frankfurt am Main}, \postcode{60438}, \state{Hesse}, \country{Germany}}}

\affil[5]{\orgdiv{División de Ciencias de la Tierra }, \orgname{Centro de Investigación Científica y de Educación Superior de Ensenada}, \orgaddress{\street{Carretera Ensenada}, \city{Tijuana}, \postcode{22860}, \state{Baja California}, \country{Mexico}}}


\abstract{Forecasting earthquake sequences remains a central challenge in seismology, particularly under non-stationary conditions. While deep learning models have shown promise, their ability to generalize across time remains poorly understood. We evaluate neural and hybrid (NN + Markov) models for short-term earthquake forecasting on a regional catalog using temporally stratified cross-validation. Models are trained on earlier portions of the catalog and evaluated on future unseen events, enabling realistic assessment of temporal generalization. We find that while these models outperform a purely Markovian model on validation data, their test performance degrades substantially in the most recent quintile. A detailed attribution analysis reveals a shift in feature relevance over time, with later data exhibiting simpler, more Markov-consistent behavior. To support interpretability, we apply Integrated Gradients, a type of explainable AI (XAI) to analyze how models rely on different input features. These results highlight the risks of overfitting to early patterns in seismicity and underscore the importance of temporally realistic benchmarks. We conclude that forecasting skill is inherently time-dependent and benefits from combining physical priors with data-driven methods.}

\keywords{Earthquake forecasting, Markov Chain, explainable AI, Neural Networks}



\maketitle

\section{Introduction}
Earthquake forecasting remains a great challenge because seismicity exhibits complex, clustered patterns that change over time. Traditional point‐process models like the Epidemic‐Type Aftershock Sequence (ETAS, \cite{Kagan1981, Kagan1987, Ogata1988, Ogata1998}) model assume a (mostly) homogeneous background rate with parameterized aftershock kernels, and have become a cornerstone of short‐term seismicity modeling. These self‐exciting models can capture the physics of aftershock sequences, but they rest on strong assumptions. For example, if the true background rate varies spatially or temporally, a homogeneous ETAS may misclassify earthquakes (mainshocks vs. aftershocks) and misestimate decay rates \cite{Stindl2022}. In practice, real earthquake catalogs often do not conform to these assumptions through nonstationary behavior (e.g., changing completeness, evolving stress). Similarly, simple first‐order Markov chain models which predict each earthquake’s region based only on the previous event provide an interpretable baseline \cite{Nava2005}, but they assume stationary transition probabilities that may not hold in a changing seismic regime. In short, both ETAS and Markov forecasts rely on statistical/physical assumptions that can break down as seismic processes evolve.
Despite these limitations, researchers have extended classical models to improve realism. For instance, Spatially Variable ETAS (SVETAS) allows ETAS parameters (like background rate) to vary across space, yielding consistently better forecasts than homogeneous ETAS \cite{Nandan2019}. They found, that in controlled experiments on California data, accounting for spatial heterogeneity leads to ``strong and statistically significant improvements in forecasting performance''\cite{Nandan2019}. Likewise, Markovian approaches have shown promise: \cite{Nava2005} modeled regional seismicity transitions as a Markov chain and reported high success rates in retrospective forecasts. More recent work even suggests that large earthquakes behave more ``Markovian'': \cite{GutierrezPeña2021} showed that only the largest-magnitude events tend to follow memoryless (Markov) transitions between regions. These findings suggest that, under some conditions, simpler memory-based models can capture a useful part of the seismic process.
In parallel, machine learning (ML) and deep learning have surged in seismology, driven by ever-growing data sets and computing power \cite{Kubo2024}. Recent reviews highlight a broad range of ML applications, from enhanced earthquake catalogs to ground-motion prediction, demonstrating great success in many areas. Spatio-temporal earthquake forecasting is a particularly active frontier. For example, \cite{DascherCousineau2023} introduced RECAST, a neural temporal point-process model based on gated recurrent units (GRUs). RECAST can ingest full earthquake catalogs and achieved forecast accuracy comparable to or better than ETAS, in tests on Southern California data, once the training catalog is large enough. Similarly, \cite{Stockman2023} developed a neural point process for the 2016–2017 Central Apennines sequence. Their model learns flexible intensity functions and was able to outperform ETAS when the catalog was highly complete: notably, the neural model remained robust with respect to missing small events, whereas ETAS performance degraded with more complete data. These studies demonstrate that ML models can capture meaningful temporal dependencies in seismicity that traditional models might miss.
Other data-driven efforts integrate physics-based features with ML. Koehler et al. (2023), for instance, used spatiotemporal maps of Gutenberg-Richter b-values as input to a deep convolutional network. Training on Japanese subduction-zone data, they achieved a binary forecasting accuracy of ~72\% (well above baseline) for predicting whether a large ($M_W>5$) quake would follow\cite{Koehler2023}. Likewise, \cite{Saad2023} proposed a multi‐modal forecasting scheme for China that combines electromagnetic (EM) and geoacoustic (GA) precursors: they extracted dozens of statistical features (via PCA and other transforms) and fed them into a machine‐learning classifier. In real‐time experiments they correctly predicted next-week large earthquakes about 70\% of the time. These hybrid approaches combining diverse geophysical signals with ML underscore the growing interest in physics-informed forecasting. At the same time, cautionary studies remind us that more complex models are not automatically better. \cite{Mignan2020} surveyed neural network efforts from 1994–2019 and found that simpler, more transparent models often match or exceed deep networks on seismic problems, given the limited and highly structured nature of earthquake catalogs. In other words, when data are scarce and features few, the apparent superiority of ``deep’’ models can evaporate; ground-truth physical insight often remains crucial.

Research gap: Crucially, most prior work has evaluated forecast models under stationary or pseudo-stationary conditions. Models are usually trained and tested on shuffled catalogs or random splits of events, effectively assuming that future seismicity follows the same patterns as the past. This ignores the reality that seismic processes evolve: tectonic loading, stress redistribution, and catalog completeness can change over time. Few studies rigorously test whether a model trained on early data can generalize to future unseen periods. As a result, high validation scores (e.g. in cross‐validation on old data) may give false confidence in a model’s ability to forecast the actual next earthquake sequence.

Goals of this study: We address this gap by focusing explicitly on temporal generalization and interpretability of neural and hybrid forecasting models. We ask: If we train on the first part of a seismic catalog, can the model accurately forecast events in later parts? To explore this, we develop a hybrid architecture that embeds a first‐order Markov structure into a neural network, allowing the model to learn both statistical transitions and complex patterns. We then perform temporal cross-validation by splitting catalogs into quintiles in time: models are trained on the earlier four quintiles and tested on the subsequent fifth, simulating a realistic prospective forecast. Finally, to peer inside the ``black box'', we apply the explainable AI (XAI) tool of integrated gradients to identify which inputs and learned features drive the model’s predictions. Similar efforts were done by \cite{Jena2023}, who use shapley additive explanation (SHAP) to asses feature importance for earthquake forecasts.

Key contributions: (1) We propose a novel neural-Markov hybrid model for earthquake sequences. (2) We establish a quintile-based temporal cross-validation framework to measure forecasting performance on truly out-of-sample future data. (3) We introduce XAI analyses, showing how to use feature importance and gradient methods to interpret what a neural model learns about past seismicity. (4) We conduct comprehensive experiments on real catalogs to rigorously compare ETAS, Markov, neural, and hybrid models under these realistic conditions.

Main findings: On random or within-sample splits, our hybrid model modestly outperforms the Markov Baseline measures with the Brier Skill Score (BSS, \cite{Brier1950}). However, when evaluated on the most recent quintile of the catalog (the ``future'' window), all models degrade sharply. Strikingly, we find a pronounced temporal asymmetry: the final quintile of seismicity behaves in a much more memoryless, Markov-consistent\footnote{We use Markov-consistent to denote, that the transition matrix for a given segment of the data corresponds well to the transition matrix calculated from the rest of the data. If this is the case, a Markov model is more likely to be a better fit than if the difference in transition matrices is larger.} fashion than the earlier data (See Figure \ref{fig:markov_deviation}). In other words, the latest period can be well predicted by a simple state-transition model, whereas the earlier period contained richer temporal dependencies. This suggests a significant non-stationarity in the catalog. The learned patterns from older earthquakes do not carry over unchanged into the new regime. This phenomenon is consistent with previous observations that only the largest earthquakes act Markovian\cite{Nava2024}, even though there are not significantly more large earthquakes in the last quintile. Importantly, this behavioral shift likely underlies the generalization gap: models trained on earlier dynamics fail to capture the changed rules of the most recent period.

Broader relevance: Our results underscore that when and how we test earthquake forecasting models is as important as what models we use. High validation scores on randomized splits can be misleading if the underlying process is nonstationary. Seismologists and data scientists should therefore favor temporally realistic evaluation (e.g. pseudo-prospective tests on held‐out future intervals). More generally, our work provides a framework for blending domain knowledge (first-order Markov structure) with deep learning and for interrogating neural forecasts with XAI. The temporal asymmetry we uncover also invites further geophysical inquiry: why does seismicity appear to ``simplify’’ in the recent period, or what made the earlier periods less Markov-consistent? Answering that could lead to better physical understanding of stress evolution and catalog completeness. For now, our study emphasizes caution: ``beware of over-optimistic forecasts’’ and shows how careful experimental design can reveal both the potentials and pitfalls of ML in seismology.

\newpage
\section{Data and Methods}

\subsection{Earthquake Catalog and State Definition}

In this work we use the dataset which was previously used in \cite{GutierrezPeña2021, Nava2024} for the approach which we use as our baseline. The catalog contains large ($M_W\geq6.5$), shallow earthquakes from 1900 to 2015 in the region surrounding the Japanese subduction zones. It is assumed to be complete above the chosen magnitude threshold.

\begin{figure}[ht]
    \centering
    \includegraphics[width=0.5\textwidth]{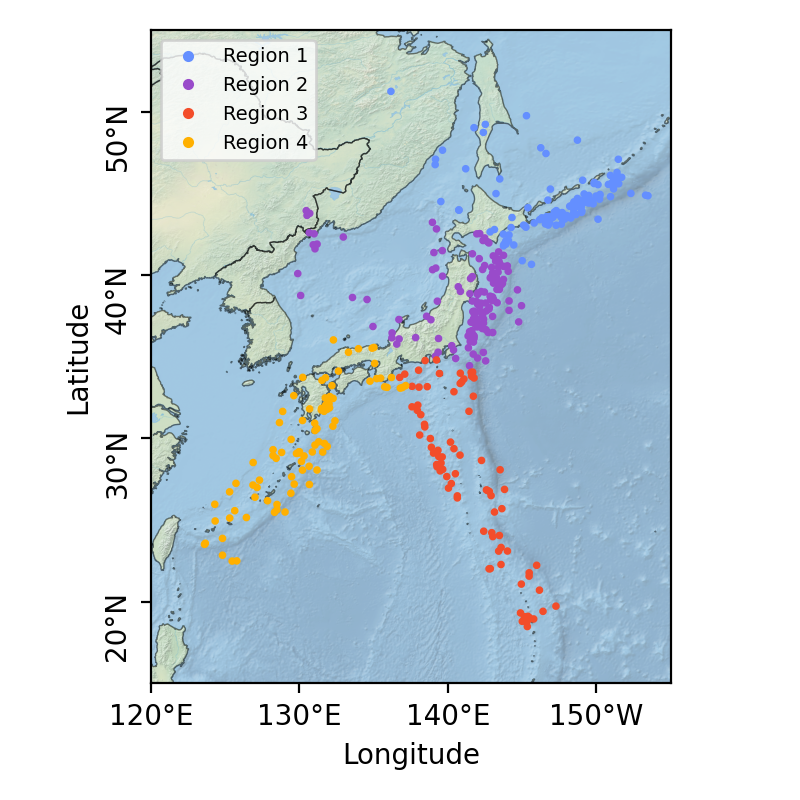} 
    \caption{Map of the study region showing all earthquakes in the catalog, color-coded by region assignment. The number of events per region is: Region 1 (120), Region 2 (152), Region 3 (85), and Region 4 (93).}
    \label{fig:map}
\end{figure}

Following their prior work, the region is divided into four tectonically defined regions, each representing a major subduction interface involving the Pacific, Philippine Sea, Okhotsk, and Amur plates (See Figure~\ref{fig:map}). These zones reflect persistent spatial clustering of large events and are designed to represent distinct seismogenic domains, each with a high potential for generating large earthquakes. At any point in time, the state of the system corresponds to the region in which the most recent large earthquake occurred, resulting in a Markov Matrix to estimate the probability of a large earthquake occurring in any of the four regions.

Since we will use not only the state but also other information related to the last event in each region, the first six events are excluded to ensure all state definitions are valid, and four more are removed at the end to allow even splitting when using a quintile base approach, resulting in 440 when using quintiles or 444 events otherwise. These are divided into five sequential, partially overlapping quintiles of 88 events each, where the final state of one quintile is used as the initial state for the next. Full independence between splits is not enforced, as sequential dependencies are intrinsic to the modeling setup and also apply to the reference Markov model.

\subsection{Input Features and Normalization}

Three sets of input features are considered:

\begin{itemize}
    \item \textbf{Last event magnitude:} The magnitude of the most recent event in each region (4 features).
    \item \textbf{Time since last event:} Elapsed time since the last event in each region, in years (4 features).
    \item \textbf{State encoding:} One-hot encoding of the current state (4 features).
\end{itemize}

Magnitudes are normalized linearly between 6.5 and 10 to yield values between 0 and $\approx0.75$ (to account for potentially bigger earthquakes). The time-since-last-event (in years) features are scaled by dividing by 10; with this rescaling, most of the values fall between 0 and 1.
However, normalization had negligible impact on model performance. 

For the pure neural network models, configurations both with and without state encoding were tested. For the hybrid model, based on the best performing pure neural network model, the state encoding is included anyways, as it was included for that model, but it is also required to extract the Markov prior.

\newpage
\subsection{Reference Model: Markov Chain}
The baseline used throughout this work is based on \cite{Nava2005, Herrera2006, GutierrezPeña2021, Nava2024}. While these works do not employ the exact same approach, it is similar enough: For our baseline, we use the 4 state $\rightarrow$ 4 state approach first used in \cite{Herrera2006} with the borders from \cite{GutierrezPeña2021}. The Markov matrix is simply calculated by counting the transitions between states, using the same set of transitions as the model this approach is compared to. Then, the resulting transition count matrix is normalized to give a transition probability matrix. This is simpler than the model brought forth in \cite{Nava2024}, but since neural networks seem to fail at outperforming even this baseline, it is sufficient.

In all cases, the transition matrix is estimated directly from the training data in that instance, ensuring that the Markov model always uses identical training data as the neural network models.\footnote{For validation, the Markov matrix is estimated for the training data. For Testing, the Markov Matrix is estimated from the training and validation data. If a model does not use validation data (while retraining a model with a fixed number of epochs), the Markov matrix is calculated only from the training data.}

\subsection{Neural and Hybrid Model Architectures}

\begin{figure}[H]
    \centering
    \includegraphics[width=0.9\textwidth]{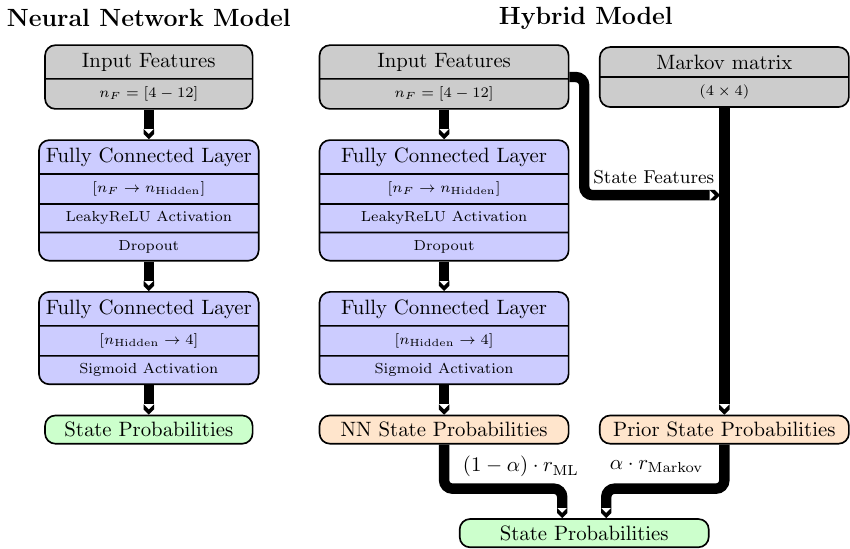}
    \caption{Architectures of the two forecasting models. Left: Pure neural network model. Right: Hybrid model combining the neural forecast with a Markov prior via a learned convex weighting.}
    \label{fig:models}
\end{figure}

We compare a compact feedforward neural network model with a hybrid variant that integrates a Markov prior into the forecasting pipeline. The architectures for both models are illustrated in Figure~\ref{fig:models}.

The pure neural network consists of a fully connected input layer (with 4 to 12 input features), a hidden layer with LeakyReLU activation, a dropout layer, and a fully connected output layer with four neurons, one for each target region, followed by a softmax activation to produce probabilistic forecasts. The model is trained to minimize the mean squared error (MSE) between predicted and observed probabilities.\footnote{While cross-entropy loss is standard for classification, it leads to overconfident predictions here. MSE, as used in the BSS, promotes calibrated probabilistic outputs, making it more appropriate for this task.} The models use the ADAM optimizer and a exponential learning rate scheduler with a $\gamma$ of 0.99 (multiplied on the learning rate each epoch).

A broad hyperparameter search using early stopping with a patience of 25 epochs, based on validation BSS (Equation \ref{eq:BSS}), was performed over the following ranges:
\begin{itemize}
    \item Hidden layer size: [4, 8, 16, 32, 64]
    \item Dropout rate: [0.0, 0.1, 0.3, 0.5]
    \item Batch size: [4, 8, 16, 32, 64]
    \item Learning rate: [$1 \times 10^{-3}$, $3 \times 10^{-4}$, $1 \times 10^{-4}$, $3 \times 10^{-5}$]
    \item Weight decay factor:\footnote{This value is multiplied with the learning rate and then used as the weight decay in the Optimizer.} [0.02, 0.5, 0.14, 0.20]
    \item Feature sets: all combinations including with/without one-hot state encoding for the neural network
\end{itemize}
\newpage
The hybrid model augments the neural forecast with a learned combination of the neural network prediction $r_{\mathrm{ML}}$ and a first-order Markov model $r_{\mathrm{Markov}}$ based on the current state:
\begin{align}
    r_{\mathrm{ML}} &= \mathrm{softmax}(\mathrm{NN}(x)) \\
    \alpha &= \mathrm{sigmoid}(\overline{\alpha}) \\
    r_{\mathrm{final}} &= \alpha \cdot r_{\mathrm{Markov}} + (1-\alpha) \cdot r_{\mathrm{ML}}
\end{align}
The Markov prior $r_{\mathrm{Markov}}$ is retrieved dynamically based on the current state. The $\overline{\alpha}$ parameter is a learnable scalar. Weight regularization was applied to all network weights except $\overline{\alpha}$, which typically converged to (rather high, and therefore potentially punished by weight decay) values between 2 and 3, corresponding to $\alpha \approx 0.88$ to $0.95$.

To ensure fair evaluation, the Markov transition matrix used at inference time in the test set is computed on the same training + validation data available to the hybrid model. For validation, the Markov model is computed only on training data, aligning with the neural network setup.

\newpage
The performances of the models are evaluated using BSS \cite{Brier1950}, defined as:

\begin{align}
    BSS &= 1 - \frac{BS_{\mathrm{ML}}}{BS_{\mathrm{MC}}} \label{eq:BSS}\\
    BS  &= \frac{1}{N} \sum_{i=1}^N \sum_{c=1}^4 (f_{ic} - r_{ic})^2
\end{align}

where $f_{ic}$ is the predicted probability for event $i$ belonging to class $c$, and $r_{ic}$ is the one-hot encoded true class, and $BS_{\mathrm{ML}}$ and $BS_{\mathrm{MC}}$ are the brier scores of the machine learning model and the Markov chain respectively. Positive BSS values indicate that the machine learning model outperforms the Markov baseline. 

\subsection{Data Splitting and Training Strategy}

\subsubsection{Fifths Experiment (Full Crossed Combinations)}

The main experiment used a fully crossed combination of the five quintiles:

\begin{itemize}
    \item 3 quintiles for training
    \item 1 quintile for validation
    \item 1 quintile for testing
\end{itemize}

For the parameter search, the testing quintile is fixed as the last quintile, while for the temporal stability analysis, we use all possible combinations: This yields $\binom{5}{3} \times \binom{2}{1} = 20$ unique train-validation-test splits. Each combination was repeated 10 times to account for stochasticity, resulting in 200 total model runs. This experiment allows evaluation of model sensitivity to the choice of training and testing intervals.

We also tested an ensemble variant restricted to using the final quintile of the data as the test set. In this approach, four models were trained, each using a different combination of three training quintiles and one validation quintile drawn from the first four quintiles of the dataset. Ensemble predictions were then formed by averaging the outputs of these four trained models.

\subsubsection{Sliding Window Event-Level Performance Evaluation}
\label{ssec:event_by_event}
To further investigate potential local variations in model performance across the full catalog, we also implemented a sliding-window evaluation procedure. The complete catalog of 444 earthquakes was used. For each evaluation, a hold-out window of size $w \in \{5, 10, 15, 20\}$ events was selected as test data. The remaining events preceding\footnote{We use the data in a loop here. If the test data consists of the first 5-20 events of the catalog, the validation set will consist of the last events of the catalog.} the test window were used for training, excluding an additional 15\% of the training data (67 events) set aside for validation and early stopping. This results in variable training set sizes depending on the test window length $w_{\mathrm{train}} \in \{372, 367, 362, 357\}$. The hold-out window was then advanced event by event across the catalog in a rolling fashion, with wrap-around applied at the end of the catalog to preserve continuity. Each unique hold-out window was evaluated five times with independent model initializations to account for stochastic training variability, resulting in $5 \times (5 + 10 + 15 + 20) = 250$ test evaluations per catalog position.

Due to the overlap of test windows, each individual event contributes to multiple BSS estimates. To obtain a smoothed estimate of event-level predictive skill, we aggregate the BSS values by summing the weighted contributions of all test windows that include a given event into an array indexed by event. After processing all windows, each element of the array is divided by the number of contributions received, yielding the average BSS attributable to each event. This averaging inherently smooths the event-level skill estimates, and thus no additional smoothing is applied.



\section{Results}
\subsection{Plain Neural Network}
We begin by testing whether the neural network can recover the Markov transition probabilities when provided only with state input, before proceeding to more complex configurations and trying to outperform the baseline.

\subsubsection{Reproducing the Markov Chain Approach}
As a first step, we verify that our shallow feedforward neural network can effectively replicate the behavior of the baseline Markov state model when given identical input information. To this end, we train models using only the one-hot encoded state information, with no magnitude or temporal inputs. Each model consists of a simple architecture trained for 2000 epochs without early stopping, to avoid validation-induced bias.\footnote{There is no validation or test set in this case. The goal here is just to show, that when using the same data, the transition probabilities from a neural network will converge to the Markov matrix.} This procedure is repeated 100 times with different random seeds to vary the initialization of values in the ML model.

After training, we evaluate each model by providing the four possible one-hot input vectors (corresponding to the four discrete states) and recording the output probabilities. These form a $4 \times 4$ transition matrix, which we compare to the original Markov model.

Note that the model parameters are different from those derived from the parameter search. To allow the network to directly replicate the structure of a Markov transition matrix, weight decay is turned off, the learning rate is set to $2\times 10^{-3}$ and no dropout is used.

\begin{figure}[ht]
    \centering
    \includegraphics[width=\textwidth]{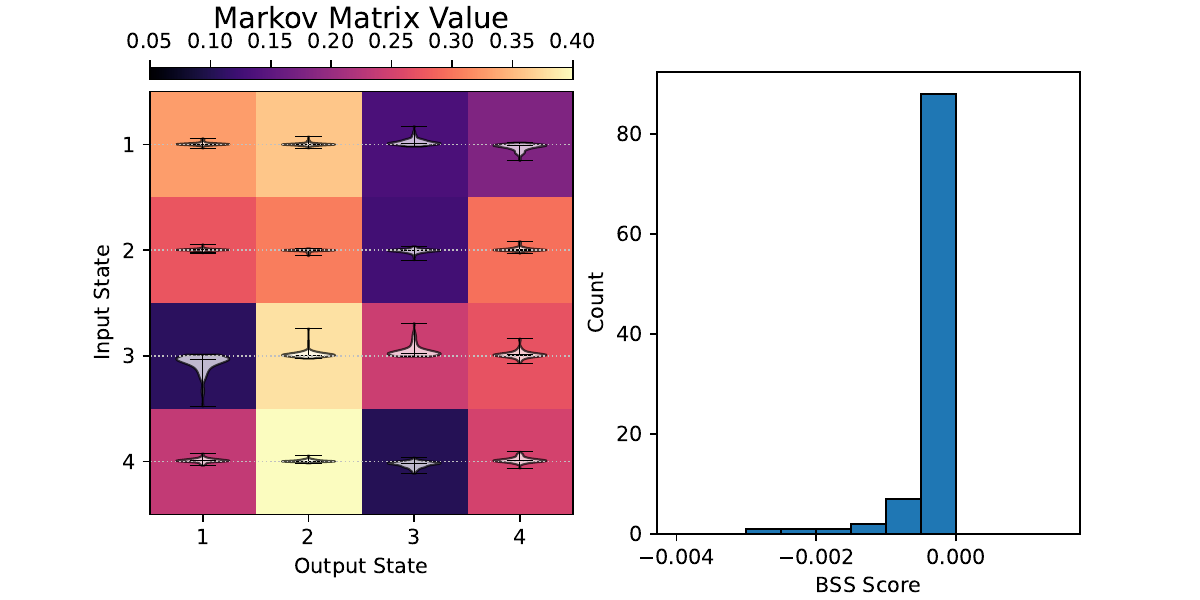} 
    \caption{Reproduction of the Markov model with a shallow neural network. \textbf{Left:} Heatmap of the Markov transition matrix. Within each cell, violin plots show the distribution of deviations from the Markov value across 100 trained networks (centered at 0). A deviation up to a cell border corresponds to a difference of $\pm 0.05 = \pm 5\%$. \textbf{Right:} Histogram of BSS achieved on the validation set by each trained network. The architecture reliably recovers the transition structure of the Markov model.}
    \label{fig:markov-reproduction}
\end{figure}

Figure~\ref{fig:markov-reproduction} (left) shows the results as a deviation heatmap: the baseline Markov matrix is shown, and overlaid violin plots visualize the distribution of deviations across the 100 trained neural networks. All entries are centered close to zero, confirming that the networks successfully reproduce the Markov transition behavior. The implied y-axis goes from $-0.05$ to $0.05$ in each colored bin, corresponding to a 5\% change for a value that falls on the bin edge.

The right panel of Figure~\ref{fig:markov-reproduction} shows a histogram of the BSS achieved by the models on the training set. Definitionally, this cannot be larger than 0, as this would outperform an optimal Markov model on a Markovian problem. The narrow spread and near-zero scores further confirm that the networks learn a behavior nearly identical to the original model under equivalent input conditions. Note that the BSS here was calculated on the training data after training, as we do not want random fluctuations of the neural network to create outperformance, which could happen as the validation or testing sets will have slightly different transition ratios.

\newpage
\subsubsection{Hyperparameter Tuning}

To identify the best-performing configuration for the plain neural network model, a comprehensive hyperparameter search was performed across six dimensions, as seen in Table~\ref{tab:hyperparameters}

\begin{table}[h!]
\centering
\caption{Hyperparameter settings for the parameter search. The one-hot encoded state is not allowed as the only Input feature, as this defaults to the Markov Baseline.}
\label{tab:hyperparameters}
\begin{tabular}{l|l}
\textbf{Hyperparameter} & \textbf{Values} \\ \hline
Batch size & 4, 8, 16, 32, 64 \\ 
Learning rate & $1 \times 10^{-3}$, $3 \times 10^{-4}$, $1 \times 10^{-4}$, $3 \times 10^{-5}$ \\ 
Weight decay factor & 0.2, 0.14, 0.05, 0.01\footnotemark \\ 
Dropout rate & 0.0, 0.1, 0.3, 0.5 \\ 
Hidden layer size & 4, 8, 16, 32 \\ 
 & (1) time since last earthquake, \\
Input feature sets & (2) magnitude of last earthquake, \\
 & (3) one-hot encoded state; used in all six allowed combinations \\
\end{tabular}
\end{table}
\footnotetext{The reported weight decay values are scaled relative to the learning rate via \texttt{weight\_decay = weight\_decay\_factor $\times$ lr.}. This formulation ensures that  the regularization strength is proportional to the learning rate, making the hyperparameter search more interpretable.}

This yields a total of $5 \times 4 \times 4 \times 4 \times 4 \times 6 = 7680$ unique hyperparameter configurations. Each configuration was trained and evaluated 10 times with different random seeds to account for stochasticity due to random weight initialization and data shuffling.

Model training used the Adam optimizer, MSE loss, and an exponential learning rate decay schedule with $\gamma = 0.99$. A maximum of 2,000 epochs was allowed per run, with early stopping (patience of 25 epochs) applied based on validation loss.\footnote{While 2000 epochs is technically possible, most models stop after 100 to 200 or between 600 to 800 epochs.} The target output was the probability distribution over the next state.

Evaluation used a fixed test set (last quintile), while the remaining four quintiles were cycled through as training and validation data using a 3:1 split. This resulted in 4 validation configurations per hyperparameter setting, and thus $7680 \times 4 \times 10 = 307,200$ individual model runs. For each configuration, the mean validation BSS across the 10 runs and across the 4 training/validation folds was used to rank performance.

\begin{figure}[H]
    \centering
    \includegraphics[width=\textwidth]{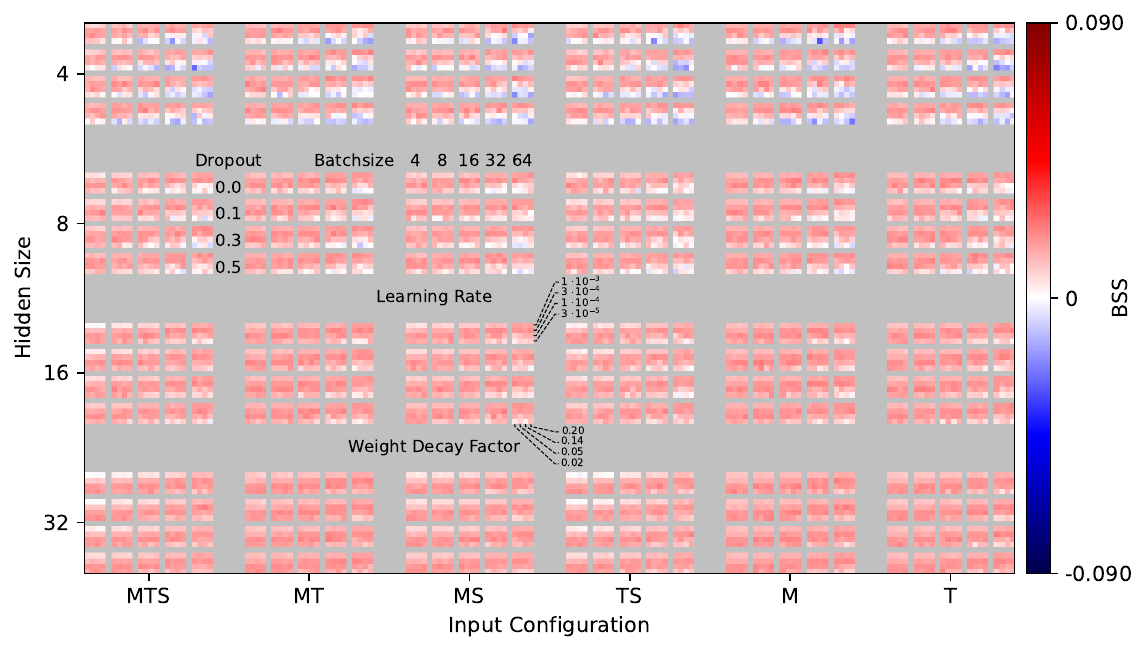} 
    \caption{Mean BSS from the full hyperparameter search, visualized as a nested heatmap. The figure is organized into 6×4 large blocks: columns correspond to input feature combinations ( Magnitude, Time and State represented with M, T and S on the X axis), and rows correspond to hidden layer sizes. Within each block, sub-rows and sub-columns represent dropout rate and batch size, respectively. Each cell inside these blocks shows a 4×4 mini-grid, where rows indicate learning rate and columns indicate weight decay. The values shown are the mean BSS across 10 runs per configuration. All axes are labeled to guide interpretation, allowing comparison across hierarchical levels of the parameter space.}
    \label{fig:PS_all}
\end{figure}

\begin{figure}[H]
    \centering
    \includegraphics[width=\textwidth]{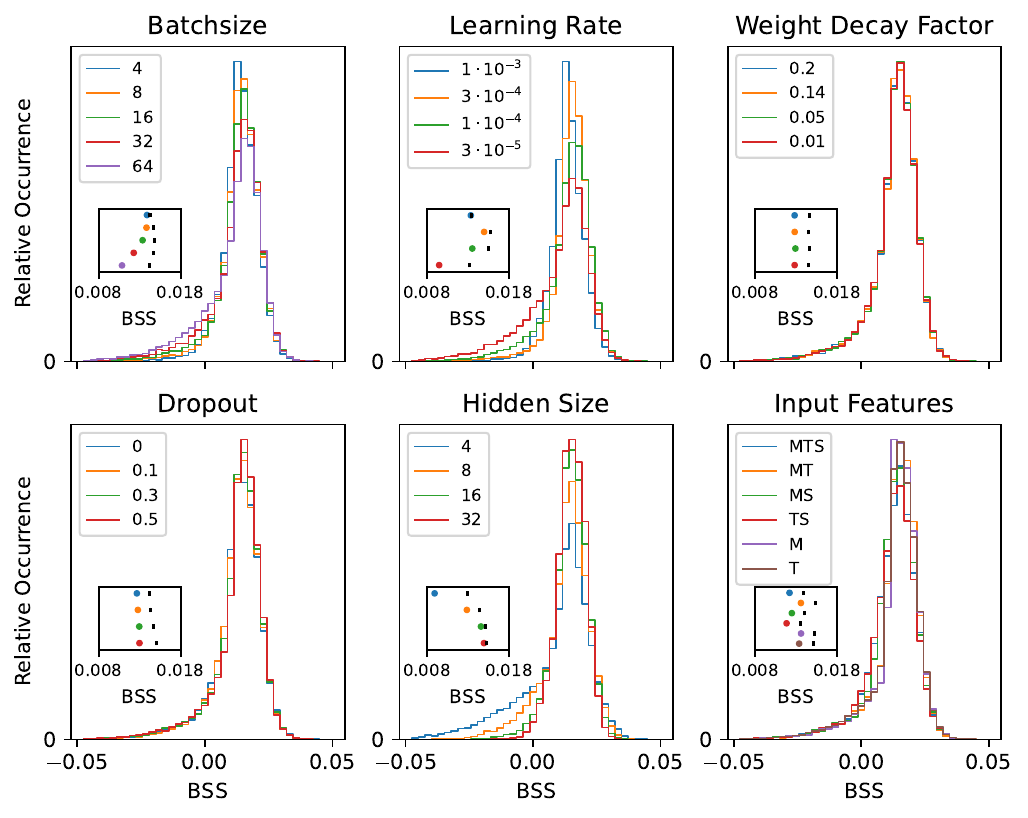} 
    \caption{Histograms of BSS for each of the six hyperparameters explored: batchsize, learning rate, weight decay factor, dropout, hidden size, input features. Each subplot displays the distribution of BSS values for that parameter, aggregated over all other settings. Inset axes show a zoomed-in view of the mean (colored dot) and median (black mark) BSS to highlight central tendencies. This visualization helps assess the relative impact and stability of each parameter choice on model performance.}
    \label{fig:PS_hist}
\end{figure}

This experiment enables identification of generalizable hyperparameter combinations that are not biased to any particular temporal segment of the first 80\% of data. The best-performing configuration is reported in Table~\ref{tab:best_hyperparams}, and performance distributions can be seen in Figures~\ref{fig:PS_all} and \ref{fig:PS_hist}.
\newpage

\newpage

To provide a thorough evaluation, three approaches for computing the final BSS against the test set are used: (1) reporting the mean validation BSS across the four train-validation splits used during hyperparameter tuning, (2) computing the BSS of an ensemble model formed by averaging the outputs of the models from each fold, and (3) retraining the best-performing architecture on the combined training and validation data (first four quintiles) and evaluating it on the held-out test quintile. For the latter, fixed-epoch training is used, with the epoch count determined by the average early stopping point of the top-performing models during tuning. These results are shown in Table~\ref{tab:bss_evaluation}.

\begin{table}
\centering
\caption{BSS of the best plain neural network model under different evaluation strategies on the test set. Each value is determined as the mean of 10 repetitions.}
\label{tab:bss_evaluation}
\begin{tabular}{llc}
\textbf{Evaluation Strategy} & \textbf{Description} & \textbf{BSS} \\
\hline
(1) Cross-Validation Average & Mean BSS across 4 validation folds & $-0.01689 \pm 0.00070$ \\
(2) Ensemble Model & Average prediction 4-fold models & $-0.01884 \pm 0.00074$ \\
(3) Retrained Model & Trained on the first 4 quintiles, fixed epochs & $-0.05306 \pm 0.00853$ \\
\end{tabular}
\end{table}

\begin{table}[h]
\centering
\caption{Best-performing hyperparameter configuration based on average validation BSS across 4 folds. The Table shows the results for both, plain neural network and hybrid model, where the daggered quantities were not included in the second parameter search, they remain the same as from the plain NN case.}
\label{tab:best_hyperparams}
\begin{tabular}{l l l}
\textbf{Hyperparameter} & \textbf{Plain NN Value} & \textbf{Hybrid Model Value}\\
\hline
Batch size & 32 & $32^\dagger$ \\
Learning rate & $3\cdot 10^{-4}$ & $1\cdot 10^{-4}$ \\
Weight decay factor & 0.05 & 0.2\\
Dropout & 0.0 & $0.0^\dagger$\\
Hidden layer size & 4 & $4^\dagger$\\
Input features & Magnitudes + States & Magnitudes + States$\mbox{}^\dagger$\\
Mean validation BSS & $0.02369 \pm 0.00214$ & $0.00426 \pm 0.00043$\\
\end{tabular}
\end{table}

\newpage
\subsection{Hybrid Approach}

To assess whether incorporating domain knowledge improves generalization, we evaluated the hybrid model, which linearly blends the output of the neural network with the Markov transition probabilities using a learnable mixing coefficient $\alpha$.

Given that the hybrid model builds directly on the architecture optimized for the plain neural network, we performed a restricted hyperparameter search, varying only the learning rate and weight decay, with the same options as for the plain model. All other hyperparameters were fixed to the best configuration found for the plain model. Each configuration was trained 10 times using early stopping (patience 25, max 2000 epochs), using the first four quintiles for training and validation, and testing on the last remaining quintile.

The learned $\alpha$ parameter, which controls the strength of the prior, typically converged to values between 0.88 and 0.95 (corresponding to $\overline{\alpha}$ between approximately 2 and 3), indicating a strong but not dominant reliance on the Markov prior. Tests with a wide range of initial values for $\overline{\alpha}$ ($0.0$ to $7.5$) consistently converged toward this range (but did not always get there), and we used an initial value of $2.5$ for the final experiments.

\begin{figure}[H]
\centering
\includegraphics[width=0.725\linewidth]{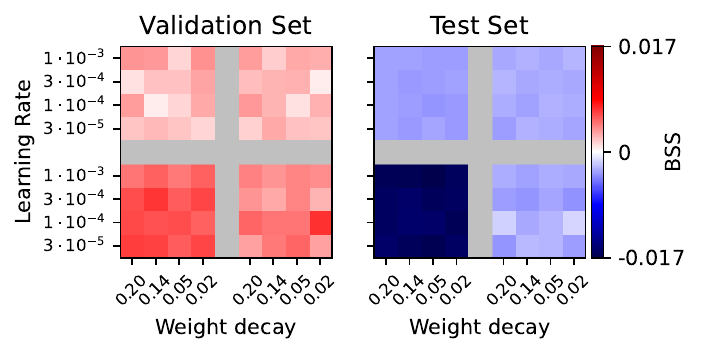}
\caption{Mean BSS for the hybrid model across learning rate (y-axis) and weight decay (x-axis), shown for each of the first four quintiles used as validation (the fifth is held out for testing). Each panel corresponds to a different validation split: $\binom{\ 1\ 2\ }{3\ 4}$. Scores are averaged over 10 random seeds per configuration. The best validation performance was achieved with a learning rate of $1\times10^{-4}$ and a weight decay of $0.20$. It is noteworthy, that the results are strongly and negatively correlated with a Pearson correlation coefficient of $-0.7043$.}
\label{fig:hybrid_paramsearch}
\end{figure}

Despite strong validation performance, test performance consistently degraded across splits (Table~\ref{tab:hybrid_bss}). A summary of validation performance across hyperparameter configurations is shown in Figure~\ref{fig:hybrid_paramsearch}. This suggests that the hybrid model, although regularized by the prior, still overfits patterns in the training and validation data that do not hold in temporally distant test data. The hybrid model also consistently outperforms the plain neural network (see Table~\ref{tab:triple_comparison}) by all three measuring standards.

These results show that even with the inclusion of an explicit prior, model performance on the last quintile is limited. This motivates the next section, where we investigate temporal inconsistencies in the dataset and their effect on model generalization.

\begin{table}[h!]
\centering
\caption{Hybrid model BSS results on validation and test sets across different data splits. The number of the split indicates which quintile was used for validation. The results shown are for 10 repetitions.}
\label{tab:hybrid_bss}
\begin{tabular}{l|cc}
\textbf{Data Split} & \textbf{Validation BSS} & \textbf{Test BSS} \\
\hline
Total   & $0.00450 \pm 0.00050$ & $-0.00591 \pm 0.00096$ \\
Split 1 & $0.00324 \pm 0.00094$ & $-0.00317 \pm 0.00014$ \\
Split 2 & $0.00347 \pm 0.00110$ & $-0.00281 \pm 0.00024$ \\
Split 3 & $0.00611 \pm 0.00050$ & $-0.01610 \pm 0.00034$ \\
Split 4 & $0.00519 \pm 0.00097$ & $-0.00158 \pm 0.00056$ \\
\end{tabular}
\end{table}

\begin{table}[h!]
\centering
\caption{BSS Comparison on the test set comparing the Markov baseline, the plain neural network and the hybrid model against one another. The first named model in the first column is considered the baseline (to interpret the sign of the BSS). The three comparison models are equivalent to those in Table~\ref{tab:bss_evaluation}, however we used 100 plain and hybrid models for improved statistical stability.}
\label{tab:triple_comparison}
\begin{tabular}{l|ccc}
\textbf{Models} & \textbf{Validation} & \textbf{Ensemble} & \textbf{Retrained} \\
\hline
Markov vs NN     & $ -0.02596 \pm 0.00044 $  & $ -0.02031 \pm 0.00039 $  & $ -0.01748 \pm 0.00060 $ \\
Markov vs Hybrid & $ -0.00639 \pm 0.00006 $  & $ -0.00004 \pm 0.00005 $  & $ -0.00013 \pm 0.00010 $ \\
NN vs Hybrid     & $ +0.01901 \pm 0.00043 $  & $ +0.01984 \pm 0.00038 $  & $ +0.01702 \pm 0.00059 $ \\
\end{tabular}
\end{table}

\newpage

\subsection{Temporal Sensitivity and Stability}
Both the plain and hybrid neural network models consistently failed to outperform the Markov baseline on the final (fifth) quintile of the dataset, regardless of validation strategy or model configuration. This underperformance is understood to stem from a shift in the underlying dynamics: the final portion of the catalog exhibits markedly more Markov-consistent behavior than earlier periods (see Section~\ref{ssec:generalization_fail} and Figure~\ref{fig:markov_deviation}). To investigate the general effects of this, we conducted two complementary analyses. First, we evaluated all combinations of training, validation, and test quintiles to assess how generalization depends on temporal location. Second, we performed a fine-grained sliding window analysis to estimate how predictive skill varies across individual events throughout the catalog. These analyses help clarify the extent and structure of temporal non-stationarity in the dataset and explain the systematic challenges faced by more flexible models in extrapolating to the final quintile.

\subsubsection{Quintiles Approach}
\begin{figure}[H]
    \centering
    \includegraphics[width=\textwidth]{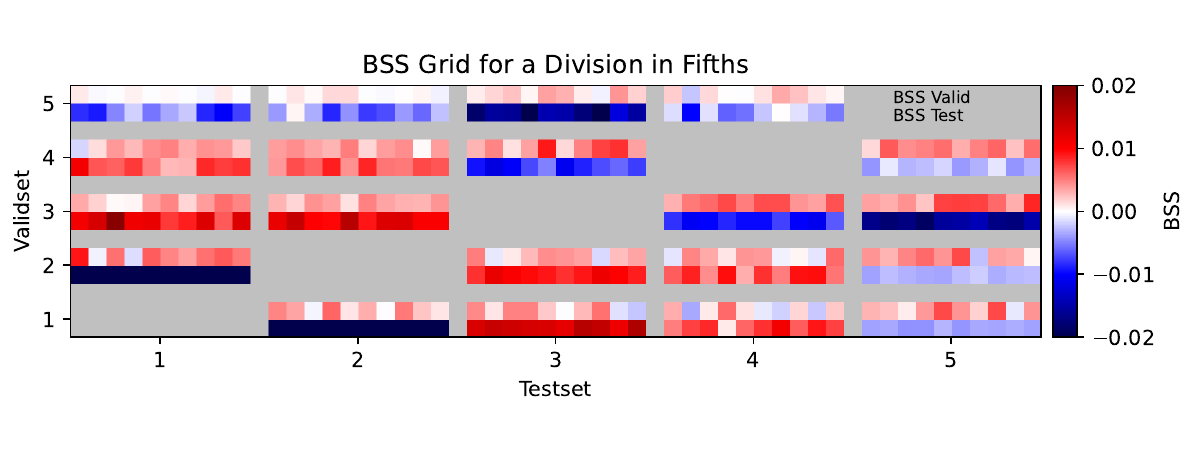}
    \caption{BSS for all combinations of training, validation, and test quintiles. Each block corresponds to a unique configuration, with the upper half showing validation BSS and the lower half showing test BSS, each repeated 10 times to account for stochastic variability. For test quintiles 1 through 4, there are always two validation splits that yield a positive mean test BSS. In contrast, Quintile~5 consistently results in negative BSS across all validation configurations, indicating poor generalization to the most recent portion of the catalog.
    }
    \label{fig:Fifths_crossvalidation}
\end{figure}

We re-evaluated the best-performing hybrid architecture across all 20 unique train-validation-test splits permitted by the five-quintile partitioning. Each configuration was run 10 times to account for stochasticity. The results, visualized in Figure~\ref{fig:Fifths_crossvalidation}, show that while individual runs within each configuration were generally stable, multiple runs were necessary to resolve variability in the BSS. Table~\ref{tab:quintile_bss_summary} summarizes BSS outcome counts and further highlights the anomalous behavior of Quintile~5 across validation and testing configurations. Importantly, test performance showed strong dependence on the specific temporal location of the test quintile. For each of the first four quintiles, at least two corresponding validation quintiles yielded a positive mean BSS relative to the Markov baseline (while two yielded a negative one). In contrast, the final quintile consistently underperformed regardless of validation set choice. Conversely, when the last quintile is used for validation, the BSS test scores of all other quintiles are also negative, likely from early stopping in non-generalizable configurations. This asymmetry highlights a pronounced temporal shift in the data distribution or dynamics near the end of the catalog that impairs generalization.

\begin{table}[h!]
\centering
\caption{Summary of negative BSS across the 20 train-validation-test combinations for each quintile. The first column counts how often the validation BSS was negative when a given quintile was used for validation. The second and third columns report how often the test BSS was negative when the quintile was used as the test or validation set, respectively. Quintile~5 shows a markedly higher number of negative outcomes, both when used as a test set and as a validation set, indicating a significant shift in the data distribution near the end of the catalog.}
\label{tab:quintile_bss_summary}
\begin{tabular}{l|c|c|c}
Quintile & $\mathrm{BSS}_\mathrm{Val} < 0$& $\mathrm{BSS}_\mathrm{Test} < 0$ (as test set)  & $\mathrm{BSS}_\mathrm{Test} < 0$ (as validation set) \\
\hline
1 & 10 & 20 & 20 \\
2 & 8  & 19 & 20 \\
3 & 0  & 20 & 20 \\
4 & 1  & 19 & 20 \\
5 & 13 & 40 & 38 \\
\end{tabular}
\end{table}

\subsubsection{Event-by-event Approach}

\begin{figure}
    \centering
    \includegraphics[width=\textwidth]{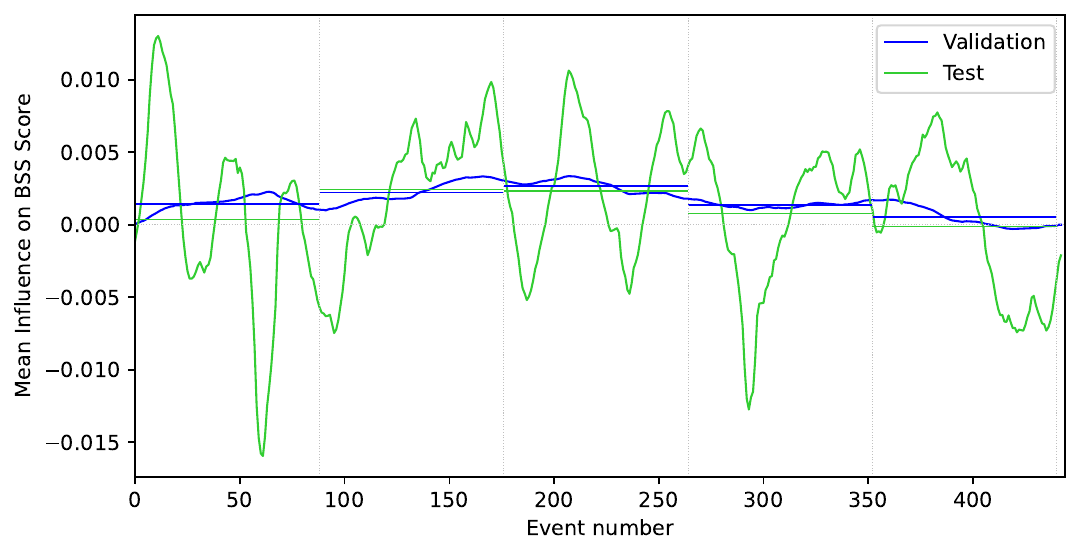}
    \caption{Smoothed event-level BSS estimates for the validation (blue) and test (green) subsets, obtained via a sliding-window evaluation with deconvolution and subsequent smoothing. Horizontal lines indicate the mean BSS for each of the five temporal quintiles. While validation BSS remains mostly positive throughout (92\% of events), test BSS displays larger variability (also due to less smoothing stemming from the smaller calculation window). The average test BSS in the last quintile is negative, in contrast to the positive quintile means elsewhere, indicating a failure of the model to generalize to the most recent events.}
    \label{fig:sliding_bss}
\end{figure}

To better understand the temporal structure of model performance, we also applied the sliding-window evaluation procedure described in Section~\ref{ssec:event_by_event} and estimated the event-level contribution to forecast skill using backpropagation, a method which inherently includes some smoothing. Figure~\ref{fig:sliding_bss} shows the resulting BSS series for both validation and test subsets.

The validation BSS remains predominantly positive throughout the catalog, with 92\% of events contributing positively to validation skill. This suggests that the model consistently fits the training data and generalizes well to held-out validation segments drawn from earlier portions of the catalog. In contrast, the test BSS series exhibits significantly more fluctuation, including extended regions of negative skill. On average, only about 60\% of events contribute positively to test BSS. Notably, a sequence of approximately 30 events in the final fifth of the catalog shows strongly negative skill (event when in the validation set), highlighting a breakdown in model generalization to the most recent portion of the data. When comparing the two lines, one should note that the validation data uses 67 events per location (repeated 20 times), while the test data uses 5-20 events per location (in steps of 5, each repeated 5 times), leading to a more smoothed validation line.

We further aggregate the BSS values across each of the five temporal quintiles. All quintiles exhibit positive mean validation BSS, while the test BSS is positive for the first four quintiles but becomes negative in the final one. This confirms the temporal asymmetry observed previously and reinforces the idea that the model struggles to extrapolate its learned representation to the most recent seismic activity.

Lastly, the Pearson correlation coefficient between the two series is approximately 0.35, suggesting a modest correspondence between the influence events have in the validation and test sets.

\newpage

\subsection{Explainable AI}
To better understand what the neural network learns from less Markov-consistent subsets of the data, we performed a focused interpretability analysis using integrated gradients (IG, \cite{Sundararajan2017}). While these results do not directly explain the model’s failure to generalize to Quintile~5, they shed light on the internal structure captured during training on earlier, more predictable portions of the catalog. IG provides a path-integrated attribution of each input feature to an output target relative to a baseline, allowing us to quantify how each input contributes to individual predictions. This offers a more fine-grained view of model behavior than global perturbation-based methods. 
The data configuration used for this analysis consists of the first 3 quintiles for training, the fourth quintile for validation and the last quintile for testing.

\begin{figure}[H]
    \centering
    \includegraphics[height=.7\textheight]{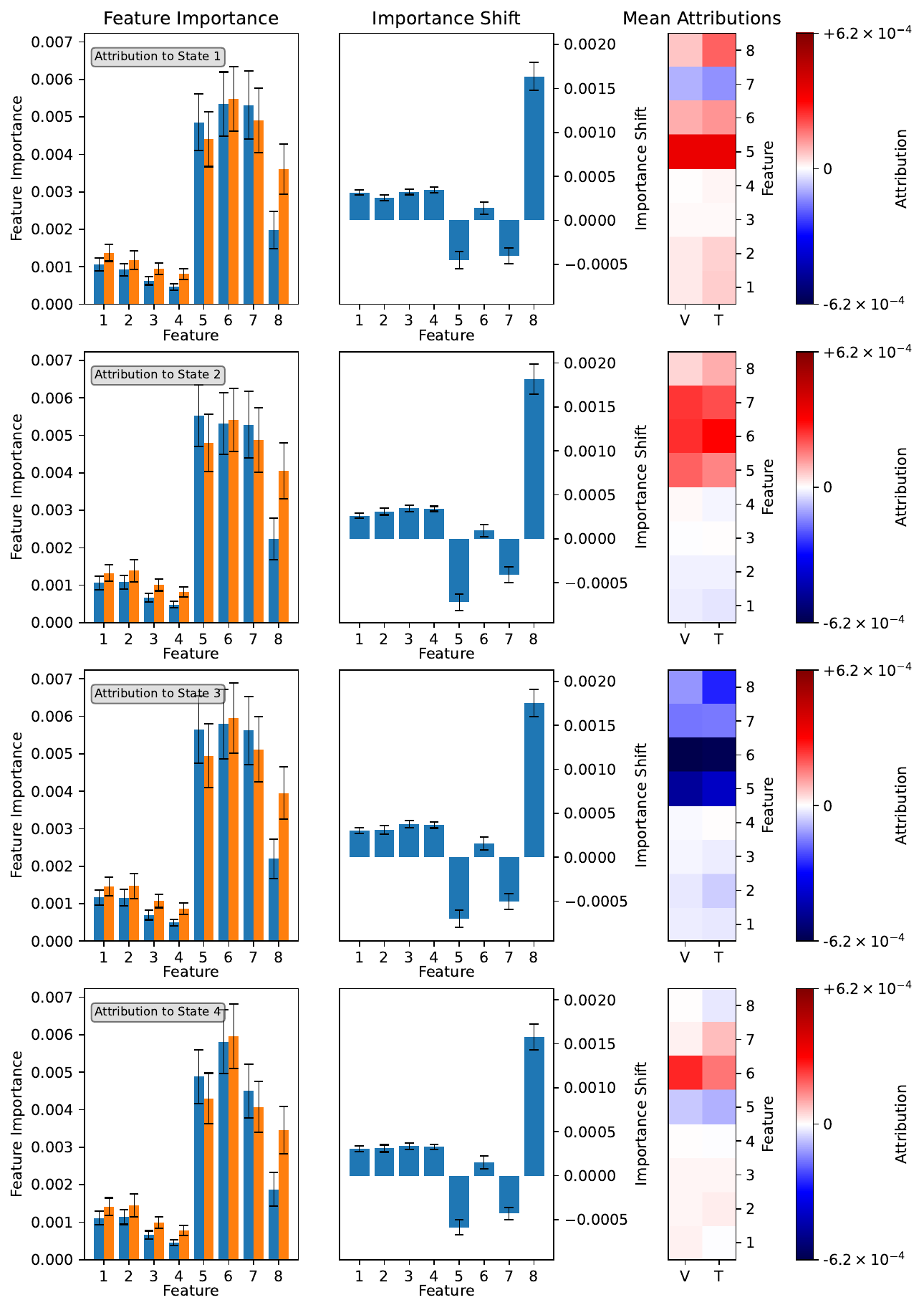}
    \caption{
    Global feature attribution for the plain neural network model using Integrated Gradients. Rows correspond to targets 1–4 from top to bottom.
    \textbf{Left}: Mean absolute IG values per feature for each output target, computed on the validation set (blue) and test set (orange). State features (5–8) dominate overall, while magnitude features (1–4) become more relevant in the test set.
    \textbf{Middle}: Difference in global attribution between test and validation sets. Features 5 and 7 show decreased importance on the test set, while feature 8 and the magnitude features increase.
    \textbf{Right}: Heatmaps of mean absolute IG values per feature and target, for validation and test sets. Rows correspond to input features 1–8, columns to validation (left) and test set (right). The dominant role of state features persists across both datasets, with moderate shifts in attribution strength.}
    \label{fig:XAI_ig_bars_global}
\end{figure}

\begin{figure}[H]
\centering
\includegraphics[height=.7\textheight]{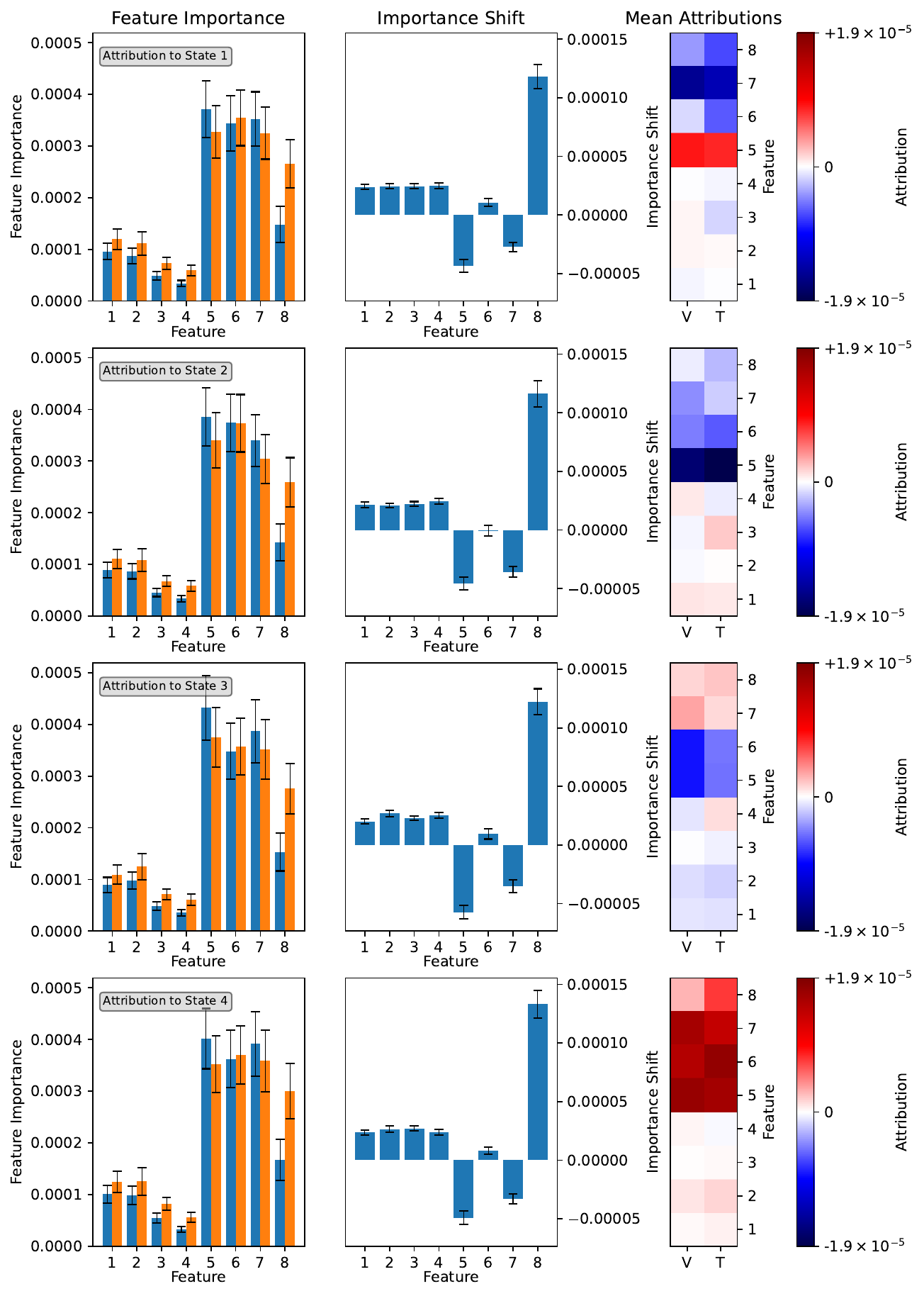}
\caption{Integrated Gradients feature attribution for the hybrid model, analogous to Figure~\ref{fig:XAI_ig_bars_global} but for the hybrid model. Left: Global feature importance for validation (blue) and test (orange) sets. Middle: Attribution shifts between validation and test sets. Right: Heatmaps of mean absolute IG values per feature and target.}
\label{fig:XAI_ig_bars_hybrid}
\end{figure}

We provide these results for both, the plain neural network (Figure~\ref{fig:XAI_ig_bars_global}) and the hybrid model (Figure~\ref{fig:XAI_ig_bars_hybrid}), although the results are very similar. The models use their respective best parameters from the parameter search and are therefore trained with eight input features: normalized magnitudes of the most recent events in each of the four regions (features 1–4), and the one-hot encoded current state (features 5–8). Magnitudes are scaled using $M_\mathrm{norm} = (M_W - 6.5)/3.5$, and state features are binary. Integrated gradients are computed for both validation and test set predictions, enabling a direct comparison of feature usage across time. These results feed into the following analysis of temporal sensitivity and model adaptation. The values discussed in this section are the mean (and standard deviation) of 100 models trained with different initial seeds.

It is generally noticeable, that the fourth target behaves differently from the others. This is likely due to its changed occurrence between the fourth and fifth quintile, which is shown in Table~\ref{tab:XAI_state_counts}.

\begin{table}[h!]
\centering
\caption{Counts of states in the validation (fourth) and test (fifth) quintile, as input and output states. Note that state 4 appears twice as often in the test set.}
\label{tab:XAI_state_counts}
\begin{tabular}{c|cc|cc}
 &  \multicolumn{2}{c}{Validation set} & \multicolumn{2}{c}{Test set} \\  
 & Input States & Output States & Input States & Output States \\ 
\hline 
State 1 & 29 & 29 & 23 & 23 \\ 
State 2 & 27 & 28 & 28 & 28 \\ 
State 3 & 24 & 23 & 21 & 21 \\ 
State 4 & 8 & 8 & 16 & 16 \\ 
\end{tabular} 

\end{table}

\vspace{1em}
\noindent\textbf{Global Feature Importance.}
We first compute the global importance of each input feature by averaging the absolute value of its IG attribution across all samples in the validation and test sets. The resulting bar plots (Figures~\ref{fig:XAI_ig_bars_global} and \ref{fig:XAI_ig_bars_hybrid}, left column) are quite similar along targets and show that the one-hot encoded state features (5–8) are generally more important across all targets. Even though the feature importance of the state features is not the same as a bar chart of these features, they match their counts closely (compare Table~\ref{tab:XAI_state_counts}).

\vspace{1em}
\noindent\textbf{Attribution Shift Between Validation and Test Sets.}
To assess temporal changes in feature usage, we calculate the difference in global attributions between validation and test sets (Figures~\ref{fig:XAI_ig_bars_global} and \ref{fig:XAI_ig_bars_hybrid}, middle column). In all cases, the magnitude features gain in importance, while the state features behave according to their prevalence (compare Table~\ref{tab:XAI_state_counts}).
These patterns suggest a temporal redistribution of feature reliance, consistent with non-stationarity or adaptation to different regimes. While higher reliance on state features (or conversely a lower reliance on magnitude features) would likely improve performance given increased Markov-consistency of the test set, the model does not recognize this and fails outperforming the baseline on the test set.

\vspace{1em}
\noindent\textbf{Feature Attribution Heatmaps.}
A summary of mean absolute IG attributions across all features and datasets is shown in the right-hand heatmaps of Figures~\ref{fig:XAI_ig_bars_global} and \ref{fig:XAI_ig_bars_hybrid}. It shows, how strongly each feature contribute to the logit for that target, on average. So while almost all features have a negative attribution for target 3 in Figure~\ref{fig:XAI_ig_bars_global}, this does not mean, that target 3 becomes less likely with an increase in each feature; other targets might decrease more, leading to a increased probability for target 3 after the final softmax function. The dominant structure remains consistent: state features receive the highest absolute attribution, (upper half of the heatmaps). While attribution shifts between validation and test sets are noticeable, they are moderate in scale, suggesting that the model retains broadly similar behavior over time, at least in aggregate.

\newpage
\section{Discussion}
While both, the plain neural network and the hybrid model outperform the Markov Chain baseline in validation, neither approach succeeds in delivering robust generalization that yields positive results on the most recent quintile of the catalog. This raises important questions about the limits of machine learning in sparse input seismic forecasting. In the following, we explore the probable cause of this failure to generalize, implications for the design and evaluation of forecasting models, and the broader value of reporting negative results in this field.

\subsection{Why Did Generalization Fail?}
\label{ssec:generalization_fail}
Although none of the models generalized successfully to the final test quintile, they consistently achieved improved forecasting skill on validation data chosen from the first four quintiles of the data. In the full cross-validation analysis, every test quintile except the last had at least two validation splits that yielded a positive BSS. This suggests that the models did learn meaningful structure from the data, but that this structure failed to remain predictive for the most recent events.

A plausible explanation is that the statistical properties of the data changed over time. Patterns learned from earlier data may no longer apply later in the catalog or may even become misleading (this can be seen in the negative correlation for BSS between validation and testsets in Figure~\ref{fig:hybrid_paramsearch}). In this view, generalization fails not because of overfitting in the usual sense, but because the target distribution itself shifts.

Regarding overfitting more generally: while the models are relatively small and the amount of training data is much larger than the number of parameters, this does not preclude the model from locking onto patterns that are only transiently predictive; it could be that there are no recognizable patterns besides Markovianity. Even with simple architectures, over-specialization to earlier regimes can occur in non-stationary settings. These findings point to the importance of explicitly evaluating temporal robustness in forecasting models.

\begin{figure}[H]
    \centering
    \includegraphics[width=\textwidth]{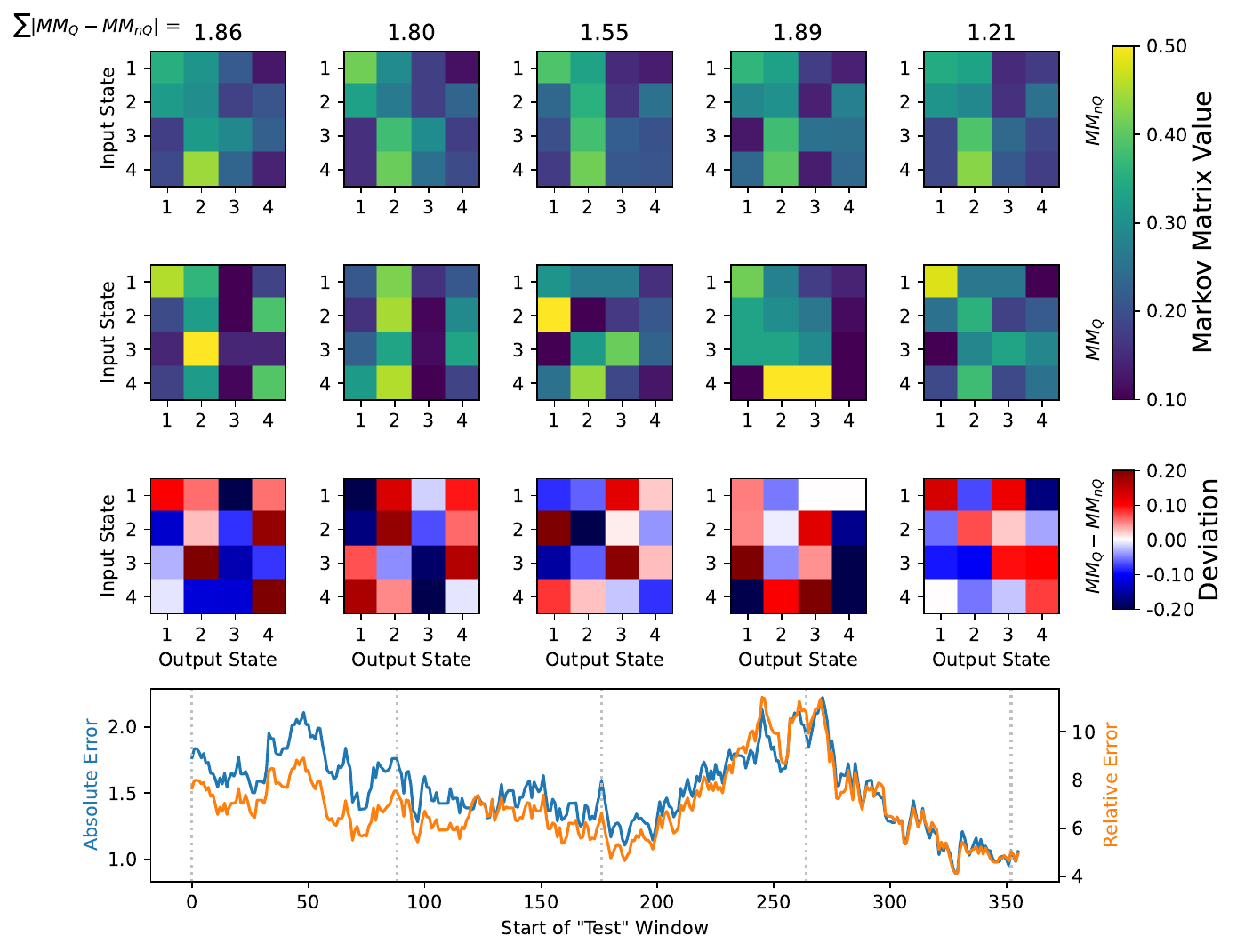}
    \caption{Comparison of Markov transition matrices across time. Each column corresponds to one of the five quintiles. The top row shows the Markov matrix computed from all other quintiles (excluding the one in the column), the middle row shows the matrix for the respective quintile, and the bottom row visualizes the absolute difference between the two. The sum of absolute differences is reported above each column. The bottom panel tracks this error in a rolling-window fashion, with a one-event step size. Both absolute and relative errors are shown. The final quintile exhibits the smallest deviation from the long-term Markov structure, suggesting that it is more Markov-consistent than the preceding data. This helps explain why models trained on earlier quintiles, which allowed greater deviation from the prior, fail to generalize when the prior becomes more predictive.}
    \label{fig:markov_deviation}
\end{figure}

To better understand the persistent degradation in test set performance, we examined the statistical structure of the data over time by analyzing changes in the Markov transition matrices across the catalog. An analysis revealed that the final quintile is substantially more consistent with the long-term Markov behavior than the earlier parts of the catalog, and a rolling-window calculation shows this even more. When comparing the transition matrix of each quintile with the matrix estimated from all other data, the fifth quintile showed the lowest deviation, both in absolute and relative terms. This suggests that the hybrid model, trained on earlier segments where deviation from the prior was larger, incorrectly learns that it can reliably depart more from the Markov baseline (the plain model which only has implicit knowledge of the Markovian nature of the problem fairs even worse, see Table~\ref{tab:triple_comparison}). However, in the final quintile, where the Markov structure is stronger, such deviations become harmful. This dynamic contributes to the poor generalization observed and reveals a broader challenge: models may not only face distribution shift, but also be systematically misled about the usefulness of their inductive biases. A representation of this Markov-consistency analysis is shown in Figure~\ref{fig:markov_deviation}.

\newpage
\subsection{Implications for Machine Learning in Sparse Forecasting Settings}
The results underscore a key risk in applying machine learning to temporally structured problems: strong performance on held-out validation data does not guarantee future generalization. In our case, models often performed well when validated on data close in time to the training set but failed when tested on more recent events. 
If only the data from the first four quintiles were available, the result of this paper would have been very different.

The hybrid model, which incorporated a probabilistic prior from a Markov process, helped stabilize learning and often improved validation scores compared to a plain neural network (Table~\ref{tab:triple_comparison}). Yet even this model failed to generalize to the final test quintile, indicating that while priors can provide useful inductive bias, they are insufficient in the face of substantial distribution shift.

These findings highlight the need for models that are either informed by additional physical variables or explicitly account for temporal change. Features such as fault-specific information, geodetic signals, or stress evolution proxies may offer the kind of conditioning necessary for generalization. Alternatively, dynamically updated models or time-aware architectures may help track evolving patterns more effectively. For sparse data models, a Markovian model might be as good as it gets.

\subsection{The Importance of Negative Results in Seismic ML}
Despite recent enthusiasm for machine learning in earthquake forecasting, the inability of our models to generalize meaningfully to the most recent data demonstrates the importance of publishing negative results. These findings help avoid overly optimistic expectations and support more realistic assessments of model utility in operational settings.

This study also illustrates the value of comprehensive evaluation. By testing models across multiple temporal splits, using ensemble strategies, fixed-epoch retraining, and event-level decomposition, we exposed shortcomings that could have remained hidden under a conventional train-validation-test split. These methods offer a practical framework for future work aiming to evaluate ML models under temporal drift.

Ultimately, documenting failure cases and carefully diagnosing their causes is essential to building more reliable models. We encourage future studies to prioritize robustness and transparency, especially when tackling problems as complex and consequential as earthquake forecasting.

\newpage
\section{Conclusion}
\subsection{Summary of Findings}
This study evaluated neural network-based classifiers for earthquake probability forecasting using sparse features derived from past events. While a neural network approach was able to outperform the performance of a Markov model during training and validation, it consistently failed to generalize to temporally distinct test data. Extensive hyperparameter tuning improved validation performance but did not resolve this gap. A hybrid model that modified Markov baseline with a learnable prior achieved higher validation skill and more stable training, yet similarly failed on the final test quintile. 
These consistent failures across model classes suggest that generalization is systematically constrained by temporal non-stationarity in the underlying data, as far as this can be determined from the limited timeseries.

\subsection{Limitations}
Several limitations should be noted, the first one being the small sample size which will affect all statistical models based on this data. The use of sparse input features limited to simple descriptors of past events likely restricts the models' ability to capture complex physical dynamics. The Markov prior, while stabilizing, is static and unable to account for possible changes in seismic regimes, even if this change is towards a more Markov-consistent state. Furthermore, the results are based on a single regional earthquake catalog without additional geophysical or contextual data. Finally, only feedforward architectures were considered; models with more explicit temporal reasoning were not explored to keep to goal of using sparse input data.

\subsection{Future Work}
Future research can build on these results while remaining within the sparse-input paradigm. One possible avenue is to refine the use of prior knowledge: dynamic extensions of the Markov model or hierarchical priors that adapt across segments of the catalog may improve robustness to non-stationarity, but the limiting factor will remain the availability of sufficiently long time series of large earthquakes.

\section{Acknowledgements}
This research is supported by the ``KI-Nachwuchswissenschaftlerinnen'' -- grant SAI 01IS20059 by the Bundesministerium für Bildung und Forschung -- BMBF. The calculations were performed on Nvidia A100 Tensor Core GPUs at the Frankfurt Institute for Advanced Studies’  GPU cluster, funded by BMBF for the project Seismologie und Artifizielle Intelligenz (SAI).
This research has made extensive use of PyTorch \cite{PyTorch2019}, numpy \cite{Numpy2020}, and matplotlib \cite{Matplotlib2007} libraries.


\bibliography{sn-bibliography.bib}

\begin{appendices}






\end{appendices}



\end{document}